\documentclass{article} 
\usepackage{main,times}
\usepackage{helvet}  
\usepackage{courier}  
\usepackage[hyphens]{url}  
\usepackage{graphicx} 
\usepackage{natbib}  
\usepackage{caption} 
\usepackage{algorithm}
\usepackage{algorithmic}
\usepackage{multirow}
\usepackage{booktabs}
\usepackage{array} 
\usepackage{makecell}
\usepackage{amsfonts}
\usepackage{amsmath}
\usepackage{amssymb}
\def\etal{\emph{et~al.}}

\def\etc.{\emph{etc}}

\usepackage{newfloat}
\usepackage{listings}
\usepackage{wrapfig}

\usepackage{hyperref}
\usepackage{url}

\title{It Hears, It Sees too: Multi-Modal LLM for Depression Detection By Integrating Visual Understanding into Audio Language Models}

\author{\textbf{Xiangyu Zhao}$^1$~\textbf{Yaling Shen}$^1$~\textbf{Yiwen Jiang}$^1$~\textbf{Zimu Wang}$^1$~\textbf{Jiahe Liu}$^1$~\textbf{Maxmartwell H Cheng}$^1$ \\
\textbf{Guilherme C Oliveira}$^1$~\textbf{Robert Desimone}$^2$~\textbf{Dominic Dwyer}$^{3}$~\textbf{Zongyuan Ge}$^1$ \\
$^1$Monash University, $^2$Massachusetts Institute of Technology, $^3$The University of Melbourne
}

\iclrfinalcopy 
\begin{document}

\maketitle

\begin{abstract}
Depression is one of the most prevalent mental health disorders globally. In recent years, multi-modal data, such as speech, video, and transcripts, has been increasingly used to develop AI-assisted depression assessment systems. Large language models have further advanced this field due to their strong language understanding and generalization capabilities. However, conventional LLMs remain text-centric and cannot process the rich non-verbal cues found in audio and visual modalities, which are critical components in mental health evaluation. While multi-modal LLMs offer a promising direction, few are tailored for psychological applications. In this study, we propose a novel multi-modal LLM framework for depression detection. Our approach augments an audio language model with visual understanding and aligns audio-visual features at the timestamp level. This fine-grained alignment improves modeling of temporal dynamics across modalities while reducing the need for extensive training data and computational resources. Experiments on the DAIC-WoZ dataset demonstrate that our model outperforms both single-modality approaches and previous multi-modal methods. Moreover, the proposed framework can be extended to incorporate additional physiological signals, paving the way for broader clinical applications beyond mental health.
\end{abstract}

\section{Introduction}
Depression has emerged as a critical concern in the field of mental health, affecting a broad population across various age groups. 
Particularly, the incidence of depression among adolescents has surged over the past decade, raising significant social and public health concerns \citep{thapar2022depression}. 
Diagnosing and treating depression often entails substantial labor and financial costs for both families and healthcare systems. With the advancement of natural language processing (NLP), increasing attention has been given to automated approaches for depression detection, reducing human intervention. Large language models (LLMs) have demonstrated remarkable capabilities across a wide array of NLP tasks \citep{naveed2023comprehensive}, which has sparked interest in their application to mental health screening \citep{hengle2024still, xu2024mental}. Despite their success, a fundamental limitation of conventional LLMs lies in their confinement to textual inputs, lacking the capacity to interpret multi-modal signals such as speech and facial expressions that are also indicative of depressive symptoms \citep{koops2023speech, krause2021facial}.

\begin{figure}[t]
  \centering
  \includegraphics[width=0.9\textwidth]{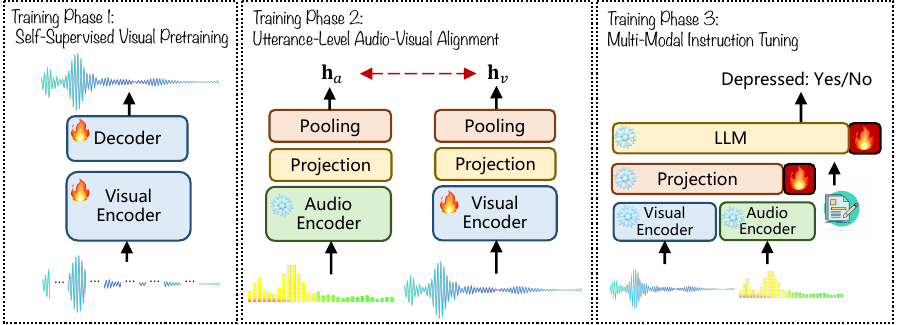}
  \caption{The training scheme of the proposed multi-modal LLM for depression detection.}
  \label{fig:scheme}
  \vspace{-15pt}
\end{figure}

Multi-modal data, including acoustic and visual cues, can significantly enhance the accuracy of depression detection. 
Prior studies have shown that individuals at high risk of depression often exhibit reduced facial expressiveness, diminished vitality, and weakened responses to external stimuli such as decreased eye contact \citep{perez2003nonverbal, waxer1974nonverbal}. Similarly, specific acoustic features, such as monotonous tone, slow speech rate, disfluency, and low vocal energy, have been linked to depressive states \citep{koops2023speech}. These behavioral signals offer valuable complementary information beyond what can be derived from text alone. 
Multi-modal large language models (MLLMs) offer an ideal solution to the integration of text and multi-modal data, which shows great promise in a lot of downstream tasks \citep{zhang2024mm}. However, current MLLMs face several limitations that hinder their application to depression detection. First, depression detection relies heavily on temporal data such as audio and video, yet most existing MLLMs are limited to static images \citep{caffagni2024revolution}. 
Furthermore, due to the relatively small size of depression-related datasets compared to standard NLP corpora, developing MLLMs for this domain demands careful consideration of model complexity to mitigate overfitting and ensure training efficiency.

To address these limitations, we propose a simple yet effective framework that adapts a multi-modal large language model for depression detection. Our method builds upon a pretrained audio language model (ALM) and augments it with visual understanding capabilities, forming a truly multi-modal system. This design leverages the shared temporal structure of audio and visual modalities, allowing for the alignment at the timestamp level. By incrementally integrating visual modules into the ALM with self-supervised visual pretraining and parameter-efficient fine-tuning (PEFT) \citep{hu2022lora}, our approach maintains the efficiency and modularity of the base model while enhancing its multi-modal capacity. This strategy also reduces the number of trainable parameters and mitigates the need for large-scale pretraining, making it efficient in data usage and computational requirements. Experiments on the public depression detection dataset, DAIC-WoZ, confirm the effectiveness of our approach, highlighting its potential for practical applications in mental health assessment.

In summary, the contributions of this work consist of the following aspects:
\begin{itemize}
    \item We develop a multi-modal large language model for depression detection based on the Qwen2-Audio \citep{chu2024qwen2} model by integrating a self-supervised vision encoder with parameter-efficient fine-tuning. To the best of our knowledge, this is the first study to propose \textbf{multi-modal depression detection using LLM across text, audio, and video modalities}; 
    \item We implement a timestamp-level alignment strategy that enables fine-grained temporal fusion across modalities. This design leverages the inherent temporal characteristics of both audio and video signals, enhancing the model’s capacity to capture subtle behavioral cues indicative of depression.
    \item We validate our approach by the comparison with single-modality methods and previous LLM-based state-of-the-art methods on the DAIC-WoZ database \citep{gratch2014distress}. The experimental results demonstrate that our approach yields superior performance at a smaller model scale (7B versus 13B), compared with pioneering multi-modal LLMs.
\end{itemize}

\section{Related Works}
\subsection{Automated Depression Detection}

Deep learning has been widely adopted for automated depression detection using speech, text, and video modalities. Earlier works focused on single modality, such as self-supervised speech models \citep{wu2023self}, hierarchical acoustic representations \citep{chen2022speechformer}, or mobile speech data \citep{kim2023automatic}. Visual features like facial expressions and eye movements have also shown promise, with methods leveraging weakly supervised learning \citep{shangguan2022dual}, gaze patterns \citep{zheng2024diagnosing}, and combined facial-gaze analysis \citep{stolicyn2022prediction}.
Recent studies have explored multi-modal fusion to capture richer cues, incorporating audio, video, and text \citep{zhang2024multimodal, shen2022automatic, xue2024fusing}. However, most rely on late fusion strategies without joint pretraining, limiting their ability to fully exploit temporal and semantic correlations across modalities.

\subsection{Large Language Models in Depression}
Large language models have been applied to depression detection due to their strong ability to model long-range dependencies in dialogue, which is an essential feature for analyzing clinical interviews. For example, Liu \etal\ \citep{liu2023chatcounselor} introduced ChatCounselor, which leverages LLMs to assess depressive symptoms and provide mental health support. Other studies have employed LLMs to analyze social media content; Hengle \etal\ \citep{hengle2024still} constructed a benchmark for depression-stress classification from online posts, while Xu \etal\ \citep{xu2024mental} used LLMs to infer depression status from various web-based sources.
Recent efforts have extended LLMs to multi-modal settings for improved diagnostic accuracy. Sadeghi \etal\ \citep{sadeghi2024harnessing} combined LLMs with facial expression analysis to estimate depression severity, and Zhang \etal\ \citep{zhang2024llms} incorporated acoustic landmarks into LLMs to build an audio-text model for depression detection. While these approaches demonstrate the potential of LLMs in mental health applications, they remain limited to textual inputs or approximations thereof (e.g., acoustic landmarks). The inability to directly process rich multi-modal signals restricts their overall effectiveness.

\subsection{Multi-Modal Large Language Models}
Integrating textual inputs with audio and visual modalities represents a major advancement in the development of generative AI. The fusion of LLMs with visual encoders has enabled impressive performance on tasks such as visual dialogue, visual question answering, and image captioning \citep{liu2023visual, zhu2023minigpt, dai2023instructblip, wang2024qwen2, lu2024deepseek}. Similarly, audio language models have emerged to jointly process speech and text. For instance, Chu \etal\ \citep{chu2024qwen2} introduced Qwen2-Audio, extending the Qwen2-7B backbone \citep{qwen2025qwen25technicalreport}, while Ding \etal\ \citep{ding2025kimi} proposed Kimi-Audio, which incorporates both discrete acoustic tokens and continuous audio embeddings into an LLM framework.
Despite their success, these models are generally not well-suited for mental health applications due to substantial domain gaps in both training data and pretraining objectives. Moreover, most vision-language models lack the capacity to handle continuous video input, further limiting their applicability to tasks such as depression detection, where temporal visual cues are crucial.

\section{Method}
\subsection{Overview of the Framework}
\begin{figure}[t]
  \centering
  \includegraphics[width=0.95\textwidth]{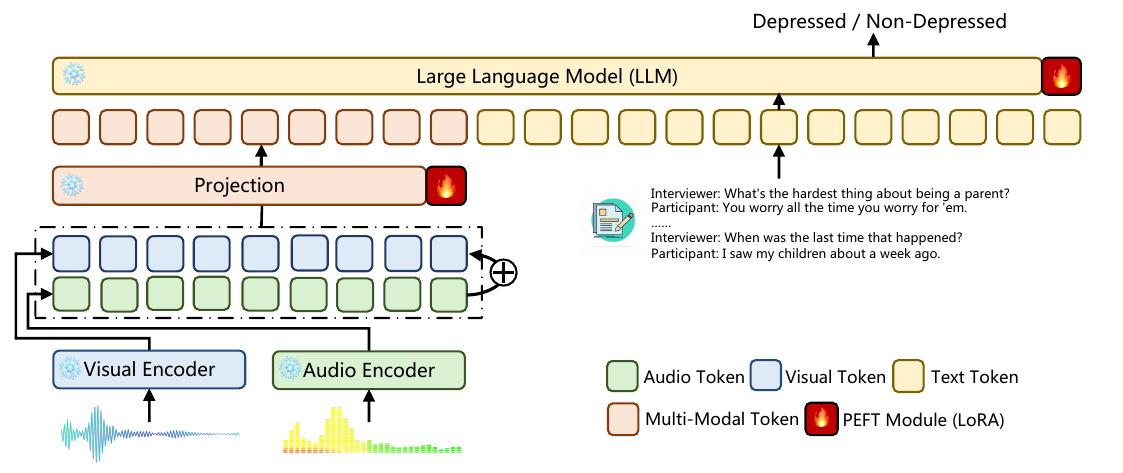}
  \caption{The framework of the proposed multi-modal large language model. The model includes an audio encoder, a visual encoder, and an LLM for detection.}
  \label{fig:framework}
  \vspace{-15pt}
\end{figure}

We propose a multi-modal large language model (MLLM) for depression detection, constructed upon a pretrained audio language model (ALM) as the backbone. As depicted in Figure~\ref{fig:framework}, the framework consists of three key components:
(1) an \textbf{audio encoder} that processes raw audio signals and extracts temporal embeddings;
(2) a \textbf{visual encoder} that receives video frames and produces visual embeddings aligned with the audio stream at the timestamp level;
(3) a \textbf{large language model} that integrates the audio-visual features along with textual inputs to perform depression classification.

The training process is divided into three sequential stages. First, the visual encoder is pretrained using a self-supervised learning strategy inspired by masked autoencoders \citep{he2022masked}, which enhances its capacity to capture rich visual representations. In the second stage, the visual encoder is fine-tuned on a contrastive alignment task designed to match visual and audio embeddings at the utterance level, thereby improving cross-modal temporal synchronization. Finally, the projection layer and LLM are trained using parameter-efficient fine-tuning (PEFT) techniques to effectively incorporate the visual modality while minimizing additional computational overhead.

\subsection{Model Components}
\subsubsection{Audio Language Model}
We adopt Qwen2-Audio \citep{chu2024qwen2} as the foundation of our framework. This model integrates Whisper-large-v3 \citep{radford2023robust} as the audio encoder and Qwen2-7B as the language model. The audio encoder processes raw waveforms resampled to 16 kHz and converts them into 128-channel Mel-spectrograms, with each frame representing a 10 ms segment. These spectrograms are subsequently downsampled via strided convolutions and average pooling, resulting in encoder outputs where each frame corresponds to a 40 ms segment of the original waveform. To ensure the universality of our method, we retain the pretrained weights of Qwen2-Audio throughout the initial stages and apply PEFT-based adaptation only in the final training phase. Notably, our framework is modular and can be extended to other audio language models, provided their audio encoders output sequences aligned with fixed temporal intervals.

\subsubsection{Visual Encoder}
The visual encoder is designed to extract visual embeddings that align temporally with the audio encoder outputs. 
To ensure architectural compatibility and ease of alignment, its design mirrors the Whisper encoder, comprising a strided convolutional embedding layer, a stack of Transformer encoder layers, and an output average pooling layer. Initially, visual features are resampled to match the temporal resolution of the audio Mel-spectrograms and are projected into the embedding space via 1D convolutions. This embedding process includes striding, reducing the temporal resolution to 20 ms per token. The resulting features are then processed by the Transformer layers and further downsampled through average pooling to match the final 40 ms resolution of the audio encoder outputs. As a result, both audio and visual embeddings are temporally synchronized, as illustrated in Figure~\ref{fig:alignment}.

\subsubsection{Audio-Visual Projection}
After obtaining audio and visual embeddings, the next step is to fuse them into a unified representation for input into the LLM. While a common fusion strategy involves concatenating modality embeddings along the sequence dimension \citep{xu2025qwen2}, this approach is suboptimal for integrating new modalities into pretrained LLMs, as it disrupts the expected sequence length and can interfere with positional encoding. To preserve compatibility with pretrained LLMs, we propose a simple yet effective fusion method—element-wise addition of audio and visual embeddings, which is illustrated in Figure~\ref{fig:framework}. This is feasible due to our explicit timestamp-level synchronization, ensuring both sequences share the same temporal structure. Moreover, our three-stage training strategy progressively aligns the modalities, enabling effective fusion without representation collapse.

\subsection{Timestamp-Synchronized Data Augmentation}
Depression corpora typically consist of participant–interviewer interviews, which present two challenges: (1) severe class imbalance, as healthy controls far outnumber depressed individuals, and (2) limited data volume, despite long session durations. To alleviate these issues, we adopt subdialogue shuffling based on \citet{wu2023self}, segmenting lengthy interviews into shorter, contiguous exchanges. This increases sample size per participant and enables flexible resampling for class balancing.

Building on \citet{wu2023self}, we enhance the method by ensuring timestamp alignment across transcript, audio, and visual modalities. Each subdialogue is constrained to start with an interviewer’s
\begin{wrapfigure}{r}{0.5\textwidth}
    \centering
    \includegraphics[width=0.5\textwidth]{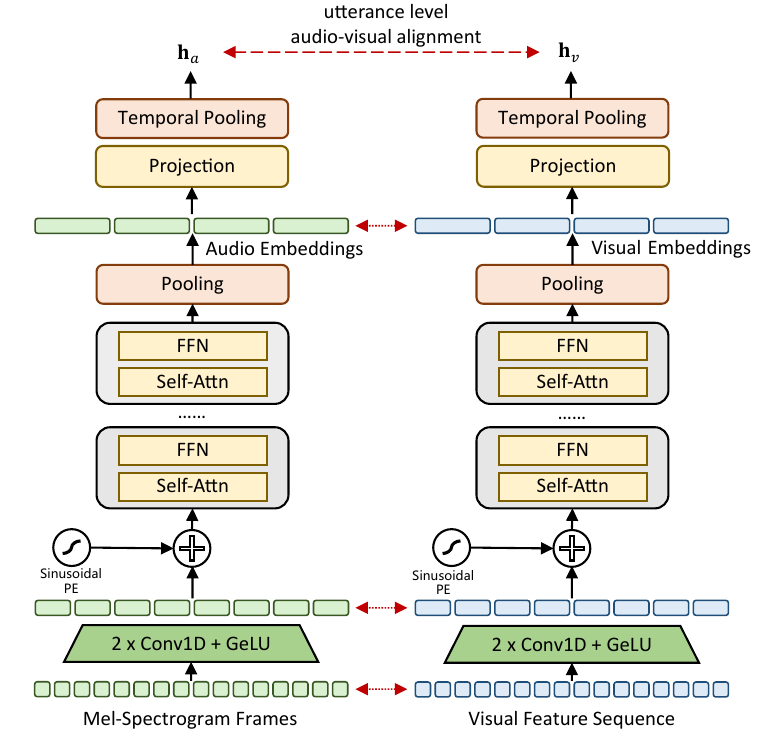}
    \caption{The scheme of utterance-level audio-visual alignment. Audio and visual inputs are strided simultaneously, ensuring synchronization on timestamps.}
    \label{fig:alignment}
\vspace{-15pt}
\end{wrapfigure}
utterance and end with the participant’s response, maintaining contextual coherence and narrowing the domain gap between LLM pretraining and depression detection. We then discard interviewer audio and corresponding visual frames, retaining only participant segments, while preserving interviewer transcripts. This choice reflects two considerations: interviewer speech carries little acoustic value for mental state assessment, yet their utterances are essential for conversational coherence. Although removing interviewer segments inevitably discards some multimodal information, the trade-off between information reduction and coherence is analyzed in Section~\ref{sec:ablation2}. Further augmentation details are provided in the Appendix.

\subsection{Training}
The training pipeline of our framework is divided into three sequential stages, as shown in Figure~\ref{fig:scheme}. The first two stages focus on training the visual encoder, while the final stage involves fine-tuning the LLM.
\subsubsection{Self-Supervised Visual Pretraining}
To enhance the visual representation capability of the encoder, we first conduct self-supervised pretraining. Instead of learning directly from raw video data, we opt to pretrain on pre-extracted visual features, as raw video files may contain sensitive content and are often unavailable in commonly used depression-related corpora. This not only addresses potential privacy concerns but also reduces computational overhead, making the approach more generalizable to other time-series modalities such as physiological signals (e.g., rPPG and ECG).

Inspired by the masked autoencoder (MAE) framework \citep{he2022masked}, we design a reconstruction task where the encoder learns to recover masked portions of the input time series. Specifically, given a sequence input $\mathbf{x} = (x_1, x_2, ..., x_T) \in \mathbb{R}^{T \times d}$, we randomly mask $K$ frames of the input and use a learnable token $x_{mask} \in \mathbb{R}^{d}$ shared across all masked frames. The indices for masked tokens are denoted as $\mathcal{M}$, and the indices for unmasked tokens are denoted as $\mathcal{V}$. Obviously $\mathcal{V} \cup \mathcal{M} = \{1, 2, \dots, T\}$. The unmasked sequence $\mathbf{x}_{in} = \{x_i | i \in \mathcal{V}\} \in \mathbb{R}^{K \times d}$ are fed to the visual encoder to acquire the latent representation $\mathbf{h} \in \mathbb{R}^{T-K \times d}$. Then, the latent representation $\mathbf{h}$ and masked frames are concatenated together to acquire the input sequence $\mathbf{z} \in \mathbb{R}^{T \times d}$, which are fed to the decoder to obtain the reconstructed input sequence $\mathbf{\hat{x}} \in \mathbb{R}^{T \times d}$. The objective is to minimize the mean squared error (MSE) between the reconstructed and original sequences within the masked regions:
\begin{equation}
    \min \frac{1}{|\mathcal{M}|}\sum_{i \in \mathcal{M}}||\hat{x}_i - x_i||^2_2
\end{equation}
This approach allows the model to capture temporal dependencies and improve robustness in downstream tasks.
\subsubsection{Utterance Level Audio-Visual Alignment}
After the visual pretraining in the first stage, the visual encoder is enabled to extract visual embeddings from input visual feature sequences for downstream tasks. However, the visual comprehension of the visual encoder is not aligned with the audio encoder. To reduce the training gap between both encoders, we design a proxy downstream task with contrastive learning to align the visual encoder with the audio encoder at the utterance level. As illustrated in Figure \ref{fig:alignment}, we add a projection layer to each encoder, respectively, and pool the outputs in the time dimension to obtain the utterance level representations. Given a mini-batch of audio outputs $\mathbf{h}_{a} \in \mathbb{R}^{N \times d}$ and visual outputs $\mathbf{h}_{v} \in \mathbb{R}^{N \times d}$, where $N$ denotes the batch size, we obtain a similarity matrix $\mathbf{Sim} = \mathbf{h}_{a}\mathbf{h}_{v}^T \in \mathbb{R}^{N \times N}$. The learning objective is to find the correct match of each audio-visual pair for utterance level audio-visual alignment:
\begin{equation}
    \min \mathcal{L}_{ce} (\mathbf{Sim} / tau, \mathbf{I}_N)
\end{equation}
where $\mathcal{L}_{ce}$ denotes cross-entropy loss, $\mathbf{I}_N$ denotes the identity matrix, and $\tau$ is the temperature parameter. 

During this stage, we freeze the entire audio encoder and the lower layers of the visual encoder to preserve the representations learned in the initial stage. Only the upper layers of the visual encoder receive gradient updates, ensuring stability and preventing catastrophic forgetting.

\subsubsection{Multi-Modal Instruction Tuning}
In the final stage, we integrate the pretrained visual encoder with the audio language model to construct a multi-modal large language model tailored for depression detection, which is illustrated in Figure \ref{fig:framework}. Since traditional LLMs are not inherently designed to process visual information, additional instruction tuning is required to adapt the model to this task. We employ Low-Rank Adaptation (LoRA) \citep{hu2022lora} to update the parameters of both the LLM and the modality projection layer. As audio and visual features have been temporally synchronized and aligned at the utterance level in previous stages, the complexity of cross-modal fusion is substantially reduced.


\subsection{Multi-Scale Sliding-Window Inference}
Since our model is trained on subdialogues rather than entire conversations, we adopt a multi-scale sliding-window inference strategy to derive a final prediction for each full conversation. This approach aggregates predictions from multiple subdialogue segments extracted at different temporal scales. Specifically, for each conversation, we generate a fixed number (200) of subdialogues at three predefined durations: 30s, 75s, and 120s. This multi-scale design ensures that each temporal resolution contributes equally to the final decision, capturing both short-term and long-term behavioral cues. The overlap between adjacent subdialogues is dynamically adjusted based on the conversation length and the total number of segments per setting.
Each time-scale configuration yields an independent conversation-level prediction, and the final prediction is determined by majority voting across the three settings.

\section{Experiments}
\subsection{Database and Implementation Details}
We utilize the DAIC-WoZ database \citep{gratch2014distress}, one of the most popular datasets for depression detection, to develop and evaluate our proposed multi-modal LLM in depression detection. The DAIC-WoZ database contains interview transcripts, speech records, and visual features from 189 participants, including healthy controls and depression cases. The golden labels of the dataset are based on PHQ-8 scores, where a PHQ-8 score higher than 10 is recognized as a depressed case. The training set contains 107 participants, 30 of whom are labeled as depressed, while the development set contains 35 participants, 12 of whom are labeled as depressed. 
Following our previous works \citep{wu2023self, zhang2024llms}, we report the evaluation results on the development set for comparison.
In addition to the training set and development set, we also evaluated our method on the test set, where 14 out of the 47 subjects are labeled as depressed.
For timestamp-synchronized data augmentation, we set the maximum length of each subdialogue to 120 seconds, generate 1,000 subdialogues per conversation with depression, which achieves a trade-off between data diversity and the risk of overfitting. The visual features generated by data augmentation are utilized for self-supervised visual pretraining and utterance level audio-visual alignment. Then the augmented transcripts, audio clips, and visual features are used for multi-modal instruction finetuning.
Our multi-modal LLM for depression detection is developed on Qwen2-Audio-7B-Instruct model. We utilize 2 NVIDIA H200 141G GPUs during training. The detailed training hyperparameters have been demonstrated in the Appendix. 
\subsection{Results}
We compare our methods with previous methods, including single-modal approaches, conventional multi-modal approaches, and multi-modal LLMs, on both the development set and test set of the DAIC-WoZ database. The detailed comparison results are illustrated in Table \ref{table:single_modal}, Table \ref{table:multi_modal}, and Table \ref{table:test_sota}, respectively. Following previous works, we adopt the F1 score for evaluation.

\paragraph{Evaluation on DAIC-WoZ Dev Set}

\begin{wraptable}{r}{5cm}
\scriptsize
\centering
\vspace{-10pt}
\begin{tabular}{clc}
\toprule
Modality               & \multicolumn{1}{c}{Models}          & F1    \\ 
\midrule
\multirow{5}{*}{Text}  & RoBERTa \citeyear{poswiata-perelkiewicz-2022-opi}  & 0.602 \\  
                       & Llama2-7B \citeyear{zhang2024llms}       & 0.578 \\
                       & Llama2-13B \citeyear{zhang2024llms}      & \textbf{0.636} \\
                       & Qwen2-7B  \citeyear{yang2024qwen2}        & 0.564 \\
                       & GPT4 \citeyear{zhang2024llms}            & 0.571 \\ 
\midrule
\multirow{5}{*}{Audio} & HuBERT \citeyear{wu2023self}         & 0.640 \\
                       & WavLM \citeyear{wu2023self}           & \textbf{0.720} \\
                       & SpeechFormer \citeyear{chen2022speechformer}   & 0.694 \\
                       & SpeechFormer++  \citeyear{chen2023speechformer++}  & 0.709 \\
                       & Whisper-v3 \citeyear{radford2023robust}     & 0.694 \\
\midrule
\multirow{2}{*}{Video} & GSM \citeyear{williamson2016detecting}    & 0.530 \\
                       & SSL + CLS    & \textbf{0.668} \\
\midrule
\multirow{3}{*}{A+T}   & AudiBERT \citeyear{toto2021audibert}  & 0.709  \\
                       & TOAT \citeyear{guo2022topic}  & 0.741  \\
                       & LSTM \citeyear{al2018detecting}    & \textbf{0.770}  \\
\midrule
\multirow{3}{*}{A+T+V} & C-CNN \citeyear{haque2018measuring}  & 0.769  \\
                       & ConvBiLSTM \citeyear{wei2022multi}  & 0.70* \\
                       & Ours w/o MS    &  0.789 \\
                       & Ours            & \textbf{0.844} \\
\bottomrule
\end{tabular}
\caption{The performance comparison of our method and other approaches on DAIC-WoZ development set. ``*" denotes that the original results are reported with 2 significant digits. ``MS" denoting the multi-scale strategy in our inference.}
\label{table:single_modal}
\vspace{-25pt}
\end{wraptable}
We present a comprehensive comparison between our proposed multi-modal LLM and previous methods on the DAIC-WoZ development set in Table \ref{table:single_modal}. Additionally, we evaluate the contribution of each individual module in our framework, including the Qwen2-7B model, the Whisper-v3 audio encoder, and a self-supervised vision encoder. Overall, our multi-modal model achieves superior classification performance on the development set of the DAIC-WoZ dataset, consistently outperforming all single-modality baselines.

Text-based models show that Llama2-13B \citep{touvron2023llama, zhang2024llms} performs best among text-only models, likely due to its larger parameter scale. Among smaller models, Qwen2-7B and Llama2-7B exhibit similar performance but fall short of the 13B variant. Interestingly, GPT-4, despite its scale and zero-shot capabilities, underperforms relative to Llama2-13B. Likewise, RoBERTa surpasses GPT-4 despite its significantly smaller size as well. A similar phenomenon has been observed in \citet{zhang2024llms}. This performance gap may be attributed to the nature of depression detection, which emphasizes representation learning over generative modeling, making encoder-based models more suitable.

Audio-based models generally outperform text-only models, suggesting that acoustic cues carry richer information for detecting depressive symptoms. 
In addition, the performance of audio models could benefit from downstream tasks such as speech recognition or emotion recognition \citep{wu2023self}. Notably, WavLM fine-tuned for emotion recognition shows superior performance, surpassing even Whisper-v3-large. This suggests that tasks closely related to depression, such as emotion recognition and ASR, provide transferable knowledge useful for this application.

For video models, our finetuned visual encoder with a classification head achieves the best performance. The main factor that could affect video-based models is the choice of visual feature sets. Since raw videos are not available at the DAIC-WoZ database, only facial feature sets, such as landmarks and action units, are available for depression detection. As the feature set could be rather redundant, the performance of video models could even deteriorate if the feature set selection is inappropriate. Self-supervised pretraining alleviates the issue significantly, as masked autoencoders are designed for images, which possess a redundant nature, and are suitable in our scenario.

Multi-modal approaches that incorporate both audio and text, or integrate all three modalities, generally outperform single-modal baselines. In particular, the inclusion of audio features often leads to significant performance improvements, highlighting the importance of acoustic information in depression detection. Compared with other multi-modal methods, our proposed framework consistently achieves superior results, demonstrating the effectiveness of timestamp-level alignment and the synergy of modality-specific encoders in capturing clinically relevant cues.

\paragraph{Comparison with Multi-Modal LLMs}
Table \ref{table:multi_modal} presents the performance comparison between our method and existing multi-modal LLMs. Together with Table \ref{table:single_modal}, the results demonstrate that incorporating audio significantly enhances the classification performance of LLMs. For instance, augmenting Llama2-13B with acoustic landmarks improves its F1 score from 0.636 to 0.695. A similar trend is observed with Qwen2-7B, where the inclusion of audio elevates the F1 score from 0.578 to 0.720. Our proposed multi-modal framework, which jointly models text, audio, and visual signals, achieves the highest F1 score of 0.789, validating the benefit of integrating visual cues alongside audio and language inputs. This underscores the advantage of leveraging complementary modalities for capturing the complex and multi-faceted nature of depressive symptoms.
\begin{wraptable}{r}{5.2cm}
\scriptsize
\centering
\vspace{-10pt}
\begin{tabular}{ccc}
\toprule
Model                         & Base Model      & F1    \\
\midrule
\multirow{4}{*}{\makecell{Acoustic LLM \\ \citep{zhang2024llms}}} & 7B  & 0.545 \\
                              & 7B-Chat & 0.500 \\ \cmidrule(l){2-3} 
                              & 13B & 0.695 \\
                              & 13B-Chat & 0.666 \\
\midrule
\multirow{2}{*}{\makecell{Qwen2-Audio\\ \citep{chu2024qwen2}}}    & 7B   & 0.650 \\
                                                                  & 7B-Instruct   & 0.720 \\
\midrule
\multirow{2}{*}{Ours w/o audio}   & 7B   & 0.617 \\
                        & 7B-Instruct   & 0.643 \\
\midrule
\multirow{2}{*}{Ours}   & 7B   & 0.709 \\
                        & 7B-Instruct   & \textbf{0.789} \\
\bottomrule
\end{tabular}
\caption{The performance comparison of our method and multi-modal LLMs on DAIC-WoZ development set. Note that for fair comparison we do not employ model ensemble or multi-scale inference.}
\label{table:multi_modal}
\vspace{-15pt}
\end{wraptable}
Notably, both our approach and Qwen2-Audio variants outperform LLMs with acoustic landmarks, despite relying on smaller language backbones (7B vs 13B). This suggests that native multi-modal architectures might be more adept at interpreting raw sensory inputs. While acoustic landmarks serve as a lightweight representation of audio, they may omit subtle prosodic or emotional cues that are preserved in the original waveforms. In contrast, models trained end-to-end on raw audio exhibit stronger modality comprehension and more effective feature fusion.

\paragraph{Evaluation on DAIC-WoZ Test Set}
In addition, since the golden labels of the DAIC-WoZ test set have been released, we compare our method with previous state-of-the-art approaches on this benchmark. The quantitative results are presented in Table \ref{table:test_sota}. It can be observed that single-modal approaches yield similar or slightly lower F1 scores on the test set compared to their performance on the development set. In contrast, a recent multi-modal approach that integrates audio, video, and textual information \citep{jung2024hique} achieves significantly better results than single-modal methods. Overall, our method outperforms both previous single-modal and multi-modal approaches on the test set, demonstrating its effectiveness and robustness.
\begin{table}[htbp]
\small
\centering
\begin{tabular}{lclc}
\toprule
Dataset                             & Models                               & Modality & F1    \\ 
\midrule
\multirow{5}{*}{DAIC-WoZ} & GloVe-CNN \citep{campbell2022speech} & Text    & 0.68* \\
                                    & TOAT \citep{guo2022topic}            & Audio    & 0.647  \\
                                    & EmoAudioNet \citep{othmani2021towards} & Audio  & 0.66*  \\
                                    & HiQuE \citep{jung2024hique}& A+T+V   & 0.79*  \\
                                    & MultiDepNet \citep{} & A+T & 0.785 \\
                                    & Ours                     & A+T+V     & \textbf{0.825}   \\
\bottomrule
\end{tabular}
\caption{The performance comparison of our method and previous approaches on DAIC-WoZ test set. ``*" denotes that the original results are reported with 2 significant digits.}
\label{table:test_sota}
\vspace{-4mm}
\end{table}

\subsection{Ablation Studies and Discussion}
In this section, we analyze the source of performance gain in our framework, including the contribution of each modality and the selection of the base model. In addition, we discuss the effectiveness of our proposed timestamp-synchronized data augmentation upon the removal of the interviewer's utterance and context length in subdialogues. The experiments are all conducted on the development set of DAIC-WoZ.
\subsubsection{The Contribution of Each Modality}
We further investigate the individual contribution of each modality within our framework. As shown in Table \ref{table:single_modal}, both audio and video modalities enhance depression detection performance. The baseline Qwen2-7B model achieves an F1 score of 0.564 using text alone. Introducing audio features leads to a substantial improvement, raising the F1 score to 0.720. Further incorporation of video features elevates the performance to 0.789. Additionally, our proposed multi-scale sliding-window strategy contributes to model performance significantly, improving the F1 score to 0.844.

An interesting observation is that the addition of audio yields a greater performance gain compared to the inclusion of video, in both instruction-tuned and pre-trained variants. This discrepancy can be attributed to two primary factors. First, as pre-extracted visual features rather than raw video data are utilized in our framework, the model may face information loss, leading to reduced expressive power. Second, our model is fundamentally built upon an audio language modeling architecture. Removing audio embeddings may disrupt the alignment mechanism across modalities, thereby compromising the model’s ability to integrate non-verbal cues effectively.

\subsubsection{The Choice of Base Model}
Since both pretrained model and instruction-tuned model are available in Qwen2-Audio families, we compare the performance of these two model variants as the base model. The results in Table \ref{table:multi_modal} indicate that the instruction-tuned model provides higher detection performance. The findings in our research are different from previous work \citep{zhang2024llms}, where instruction tuning leads to significant performance deterioration compared with the pretrained model. The reasons for the inconsistency could be the difference in instruction tuning in general LLMs and audio language models. Depression detection involves the analysis of both audio and text; a similar task has been used to finetune the model in instruction tuning. Thus, the instruction-tuned model could be better at the audio analysis task. 

\subsubsection{The Effect of Context Length and Interviewer Utterance Removal}
\label{sec:ablation2}
\begin{wrapfigure}{r}{0.5\textwidth}
\vspace{-5mm}
  \includegraphics[width=0.5\textwidth]{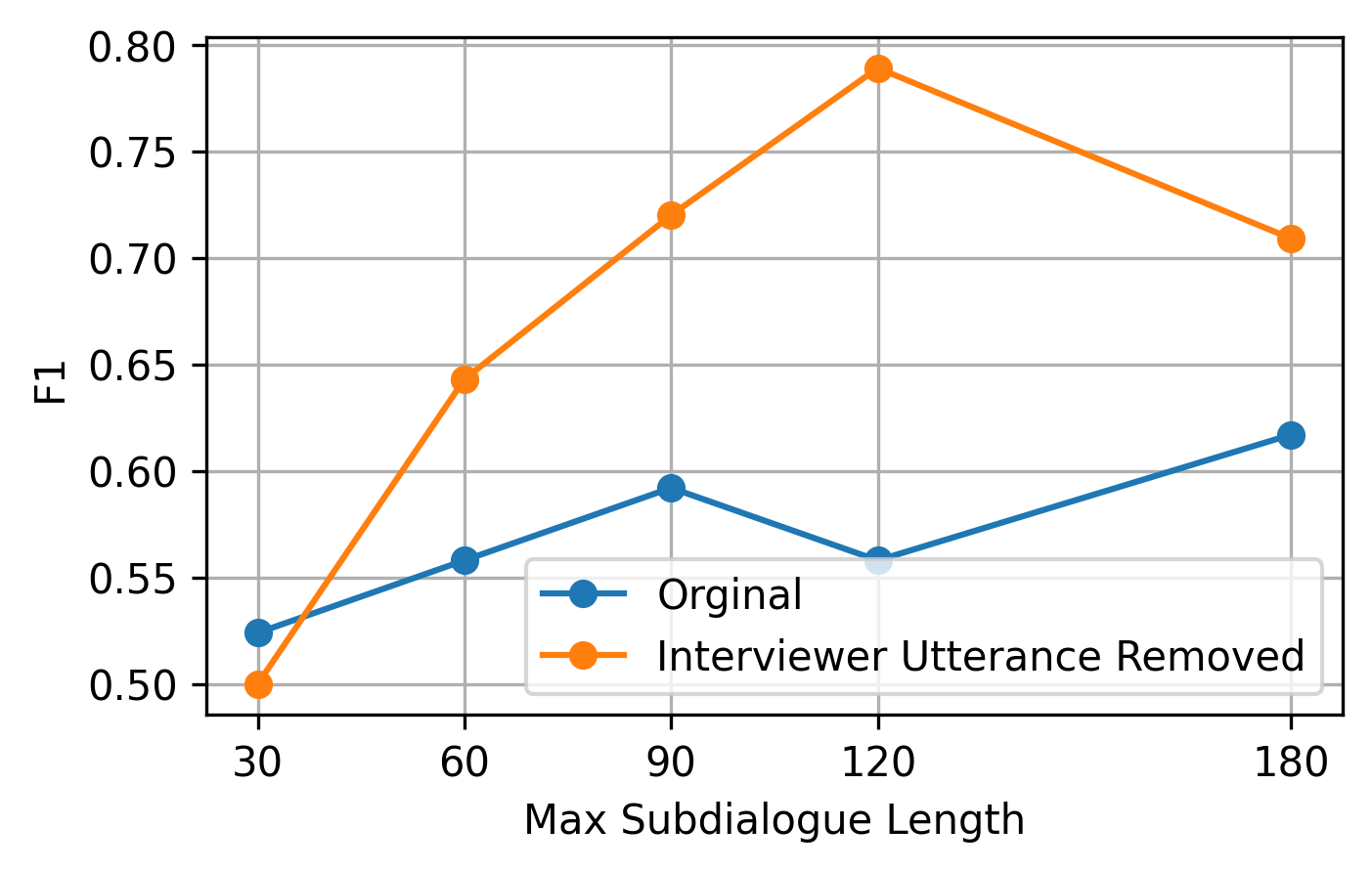}
  \caption{The depression detection performance on subdialogue length with or without interviewer's utterances.}
  \label{fig:ablation}
  \vspace{-5mm}
\end{wrapfigure}
The length of subdialogues plays a crucial role in our framework, as longer contexts generally provide richer cues for depression detection. However, longer subdialogues do not necessarily improve the performance for detection, as the audio records for the interviewer do not contribute to the decision, but even interfere with the depression detection. To address this constraint, we propose to remove the interviewer’s utterances during data augmentation, allowing more content from the participant to be retained within the fixed audio window. While this enhances the availability of participant-specific acoustic cues, it also results in the loss of visual information associated with the removed segments.
To explore this trade-off, we conduct an ablation study under varying subdialogue lengths, 
as shown in Figure \ref{fig:ablation}. When the maximum subdialogue length is constrained to 30 seconds, removing the interviewer’s speech leads to degraded performance. 
In this setting, the entire subdialogue can be encoded without truncation, and discarding the interviewer’s turns causes unnecessary loss of visual cues, thus impairing multi-modal inference. In contrast, as the subdialogue length increases beyond the model’s audio capacity, the removal of interviewer utterances proves beneficial. By prioritizing participant speech within the fixed input window, the model gains access to more relevant acoustic information, leading to improved detection accuracy. However, when the context length becomes excessively long, the performance gain diminishes. This is likely due to reduced dialogue diversity and increased risk of overfitting, as longer subdialogues tend to be less variable.

\section{Conclusion}
In this study, we propose a multi-modal large language model for depression detection, built upon audio-based language models and augmented with visual understanding capabilities. Experiments on the DAIC-WoZ dataset demonstrate the superiority of our framework over existing multi-modal LLMs. To our knowledge, this is the first work to develop a multi-modal LLM for depression detection that simultaneously integrates textual, audio, and visual modalities. We further provide detailed analyses of how model design and data augmentation strategies affect performance. Overall, our method offers an effective solution for adapting multi-modal LLMs to mental health applications, with potential for broader extension to other domains.

\section*{Ethics Statement}
This study is conducted using the DAIC-WoZ database, a publicly available resource accessible to qualified researchers upon request. All data collection procedures for this dataset were carried out with informed consent from participants, and the data have been fully anonymized to protect individual privacy. We have obtained proper authorization by signing the DAIC-WoZ End-User License Agreement and strictly adhere to its terms of use.
Our model is built upon the Qwen2-Audio architecture, and all research activities related to it comply with the Apache-2.0 license under which the model is released.
While our method achieves state-of-the-art performance on the DAIC-WoZ benchmark, it is intended for research purposes only and should not be used for clinical diagnosis, treatment, or intervention of depression. We further acknowledge that, like many large language models, our framework may be vulnerable to hallucinations, harmful outputs, or systemic biases. We disclaim responsibility for any misuse, misinterpretation, or unintended consequences resulting from the deployment of this model outside its intended research context.

\bibliographystyle{main}
\bibliography{references}

@article{thapar2022depression,
  title={Depression in young people},
  author={Thapar, Anita and Eyre, Olga and Patel, Vikram and Brent, David},
  journal={The Lancet},
  volume={400},
  number={10352},
  pages={617--631},
  year={2022},
  publisher={Elsevier}
}

@article{koops2023speech,
  title={Speech as a biomarker for depression},
  author={Koops, Sanne and Brederoo, Sanne G and de Boer, Janna N and Nadema, Femke G and Voppel, Alban E and Sommer, Iris E},
  journal={CNS \& Neurological Disorders-Drug Targets-CNS \& Neurological Disorders)},
  volume={22},
  number={2},
  pages={152--160},
  year={2023},
  publisher={Bentham Science Publishers direct}
}

@inproceedings{wu2023self,
  title={Self-supervised representations in speech-based depression detection},
  author={Wu, Wen and Zhang, Chao and Woodland, Philip C},
  booktitle={ICASSP 2023-2023 IEEE International Conference on Acoustics, Speech and Signal Processing (ICASSP)},
  pages={1--5},
  year={2023},
  organization={IEEE}
}

@inproceedings{chen2022speechformer,
  title={SpeechFormer: A Hierarchical Efficient Framework Incorporating the Characteristics of Speech},
  author={Chen, Weidong and Xing, Xiaofen and Xu, Xiangmin and Pang, Jianxin and Du, Lan},
  booktitle={Proc. Interspeech 2022},
  pages={346--350},
  year={2022}
}

@article{kim2023automatic,
  title={Automatic depression detection using smartphone-based text-dependent speech signals: deep convolutional neural network approach},
  author={Kim, Ah Young and Jang, Eun Hye and Lee, Seung-Hwan and Choi, Kwang-Yeon and Park, Jeon Gue and Shin, Hyun-Chool},
  journal={Journal of medical Internet research},
  volume={25},
  pages={e34474},
  year={2023},
  publisher={JMIR Publications Toronto, Canada}
}

@article{shangguan2022dual,
  title={Dual-stream multiple instance learning for depression detection with facial expression videos},
  author={Shangguan, Zixuan and Liu, Zhenyu and Li, Gang and Chen, Qiongqiong and Ding, Zhijie and Hu, Bin},
  journal={IEEE Transactions on Neural Systems and Rehabilitation Engineering},
  volume={31},
  pages={554--563},
  year={2022},
  publisher={IEEE}
}

@article{zheng2024diagnosing,
  title={Diagnosing and tracking depression based on eye movement in response to virtual reality},
  author={Zheng, Zhiguo and Liang, Lijuan and Luo, Xiong and Chen, Jie and Lin, Meirong and Wang, Guanjun and Xue, Chenyang},
  journal={Frontiers in Psychiatry},
  volume={15},
  pages={1280935},
  year={2024},
  publisher={Frontiers Media SA}
}

@article{stolicyn2022prediction,
  title={Prediction of depression symptoms in individual subjects with face and eye movement tracking},
  author={Stolicyn, Aleks and Steele, J Douglas and Seri{\`e}s, Peggy},
  journal={Psychological medicine},
  volume={52},
  number={9},
  pages={1784--1792},
  year={2022},
  publisher={Cambridge University Press}
}

@article{zhang2024multimodal,
  title={Multimodal sensing for depression risk detection: Integrating audio, video, and text data},
  author={Zhang, Zhenwei and Zhang, Shengming and Ni, Dong and Wei, Zhaoguo and Yang, Kongjun and Jin, Shan and Huang, Gan and Liang, Zhen and Zhang, Li and Li, Linling and others},
  journal={Sensors},
  volume={24},
  number={12},
  pages={3714},
  year={2024},
  publisher={MDPI}
}

@article{sadeghi2024harnessing,
  title={Harnessing multimodal approaches for depression detection using large language models and facial expressions},
  author={Sadeghi, Misha and Richer, Robert and Egger, Bernhard and Schindler-Gmelch, Lena and Rupp, Lydia Helene and Rahimi, Farnaz and Berking, Matthias and Eskofier, Bjoern M},
  journal={npj Mental Health Research},
  volume={3},
  number={1},
  pages={66},
  year={2024},
  publisher={Nature Publishing Group UK London}
}

@inproceedings{shen2022automatic,
  title={Automatic depression detection: An emotional audio-textual corpus and a gru/bilstm-based model},
  author={Shen, Ying and Yang, Huiyu and Lin, Lin},
  booktitle={ICASSP 2022-2022 IEEE International Conference on Acoustics, Speech and Signal Processing (ICASSP)},
  pages={6247--6251},
  year={2022},
  organization={IEEE}
}

@inproceedings{xue2024fusing,
  title={Fusing multi-level features from audio and contextual sentence embedding from text for interview-based depression detection},
  author={Xue, Junqi and Qin, Ruihan and Zhou, Xinxu and Liu, Honghai and Zhang, Min and Zhang, Zhiguo},
  booktitle={ICASSP 2024-2024 IEEE International Conference on Acoustics, Speech and Signal Processing (ICASSP)},
  pages={6790--6794},
  year={2024},
  organization={IEEE}
}

@article{liu2023chatcounselor,
  title={Chatcounselor: A large language models for mental health support},
  author={Liu, June M and Li, Donghao and Cao, He and Ren, Tianhe and Liao, Zeyi and Wu, Jiamin},
  journal={arXiv preprint arXiv:2309.15461},
  year={2023}
}

@inproceedings{hengle2024still,
  title={Still Not Quite There! Evaluating Large Language Models for Comorbid Mental Health Diagnosis},
  author={Hengle, Amey and Kulkarni, Atharva and Patankar, Shantanu and Chandrasekaran, Madhumitha and D’silva, Sneha and Jacob, Jemima and Gupta, Rashmi},
  booktitle={Proceedings of the 2024 Conference on Empirical Methods in Natural Language Processing},
  pages={16698--16721},
  year={2024}
}

@article{xu2024mental,
  title={Mental-llm: Leveraging large language models for mental health prediction via online text data},
  author={Xu, Xuhai and Yao, Bingsheng and Dong, Yuanzhe and Gabriel, Saadia and Yu, Hong and Hendler, James and Ghassemi, Marzyeh and Dey, Anind K and Wang, Dakuo},
  journal={Proceedings of the ACM on Interactive, Mobile, Wearable and Ubiquitous Technologies},
  volume={8},
  number={1},
  pages={1--32},
  year={2024},
  publisher={ACM New York, NY, USA}
}

@inproceedings{zhang2024llms,
  title={When LLMs Meets Acoustic Landmarks: An Efficient Approach to Integrate Speech into Large Language Models for Depression Detection},
  author={Zhang, Xiangyu and Liu, Hexin and Xu, Kaishuai and Zhang, Qiquan and Liu, Daijiao and Ahmed, Beena and Epps, Julien},
  booktitle={Proceedings of the 2024 Conference on Empirical Methods in Natural Language Processing},
  pages={146--158},
  year={2024}
}

@article{liu2023visual,
  title={Visual instruction tuning},
  author={Liu, Haotian and Li, Chunyuan and Wu, Qingyang and Lee, Yong Jae},
  journal={Advances in neural information processing systems},
  volume={36},
  pages={34892--34916},
  year={2023}
}

@article{zhu2023minigpt,
  title={Minigpt-4: Enhancing vision-language understanding with advanced large language models},
  author={Zhu, Deyao and Chen, Jun and Shen, Xiaoqian and Li, Xiang and Elhoseiny, Mohamed},
  journal={arXiv preprint arXiv:2304.10592},
  year={2023}
}

@inproceedings{
dai2023instructblip,
title={Instruct{BLIP}: Towards General-purpose Vision-Language Models with Instruction Tuning},
author={Wenliang Dai and Junnan Li and Dongxu Li and Anthony Tiong and Junqi Zhao and Weisheng Wang and Boyang Li and Pascale Fung and Steven Hoi},
booktitle={Thirty-seventh Conference on Neural Information Processing Systems},
year={2023},
url={https://openreview.net/forum?id=vvoWPYqZJA}
}

@article{wang2024qwen2,
  title={Qwen2-vl: Enhancing vision-language model's perception of the world at any resolution},
  author={Wang, Peng and Bai, Shuai and Tan, Sinan and Wang, Shijie and Fan, Zhihao and Bai, Jinze and Chen, Keqin and Liu, Xuejing and Wang, Jialin and Ge, Wenbin and others},
  journal={arXiv preprint arXiv:2409.12191},
  year={2024}
}

@article{lu2024deepseek,
  title={Deepseek-vl: towards real-world vision-language understanding},
  author={Lu, Haoyu and Liu, Wen and Zhang, Bo and Wang, Bingxuan and Dong, Kai and Liu, Bo and Sun, Jingxiang and Ren, Tongzheng and Li, Zhuoshu and Yang, Hao and others},
  journal={arXiv preprint arXiv:2403.05525},
  year={2024}
}

@article{chu2024qwen2,
  title={Qwen2-audio technical report},
  author={Chu, Yunfei and Xu, Jin and Yang, Qian and Wei, Haojie and Wei, Xipin and Guo, Zhifang and Leng, Yichong and Lv, Yuanjun and He, Jinzheng and Lin, Junyang and others},
  journal={arXiv preprint arXiv:2407.10759},
  year={2024}
}

@misc{qwen2025qwen25technicalreport,
      title={Qwen2.5 Technical Report}, 
      author={Qwen and : and An Yang and Baosong Yang and Beichen Zhang and Binyuan Hui and Bo Zheng and Bowen Yu and Chengyuan Li and Dayiheng Liu and Fei Huang and Haoran Wei and Huan Lin and Jian Yang and Jianhong Tu and Jianwei Zhang and Jianxin Yang and Jiaxi Yang and Jingren Zhou and Junyang Lin and Kai Dang and Keming Lu and Keqin Bao and Kexin Yang and Le Yu and Mei Li and Mingfeng Xue and Pei Zhang and Qin Zhu and Rui Men and Runji Lin and Tianhao Li and Tianyi Tang and Tingyu Xia and Xingzhang Ren and Xuancheng Ren and Yang Fan and Yang Su and Yichang Zhang and Yu Wan and Yuqiong Liu and Zeyu Cui and Zhenru Zhang and Zihan Qiu},
      year={2025},
      eprint={2412.15115},
      archivePrefix={arXiv},
      primaryClass={cs.CL},
      url={https://arxiv.org/abs/2412.15115}, 
}

@article{ding2025kimi,
  title={Kimi-Audio Technical Report},
  author={Ding, Ding and Ju, Zeqian and Leng, Yichong and Liu, Songxiang and Liu, Tong and Shang, Zeyu and Shen, Kai and Song, Wei and Tan, Xu and Tang, Heyi and others},
  journal={arXiv preprint arXiv:2504.18425},
  year={2025}
}

@article{naveed2023comprehensive,
  title={A comprehensive overview of large language models},
  author={Naveed, Humza and Khan, Asad Ullah and Qiu, Shi and Saqib, Muhammad and Anwar, Saeed and Usman, Muhammad and Akhtar, Naveed and Barnes, Nick and Mian, Ajmal},
  journal={arXiv preprint arXiv:2307.06435},
  year={2023}
}

@article{krause2021facial,
  title={Facial emotion recognition in major depressive disorder: A meta-analytic review},
  author={Krause, Fernando C and Linardatos, Eftihia and Fresco, David M and Moore, Michael T},
  journal={Journal of affective disorders},
  volume={293},
  pages={320--328},
  year={2021},
  publisher={Elsevier}
}

@article{perez2003nonverbal,
  title={Nonverbal social skills and psychopathology},
  author={Perez, John E and Riggio, Ronald E},
  journal={Nonverbal behavior in clinical settings},
  pages={17--44},
  year={2003},
  publisher={Oxford University Press New York}
}

@article{waxer1974nonverbal,
  title={Nonverbal cues for depression.},
  author={Waxer, Peter},
  journal={Journal of Abnormal Psychology},
  volume={83},
  number={3},
  pages={319},
  year={1974},
  publisher={American Psychological Association}
}

@inproceedings{gratch2014distress,
  title={The distress analysis interview corpus of human and computer interviews.},
  author={Gratch, Jonathan and Artstein, Ron and Lucas, Gale M and Stratou, Giota and Scherer, Stefan and Nazarian, Angela and Wood, Rachel and Boberg, Jill and DeVault, David and Marsella, Stacy and others},
  booktitle={LREC},
  volume={14},
  pages={3123--3128},
  year={2014},
  organization={Reykjavik}
}

@inproceedings{he2022masked,
  title={Masked autoencoders are scalable vision learners},
  author={He, Kaiming and Chen, Xinlei and Xie, Saining and Li, Yanghao and Doll{\'a}r, Piotr and Girshick, Ross},
  booktitle={Proceedings of the IEEE/CVF conference on computer vision and pattern recognition},
  pages={16000--16009},
  year={2022}
}

@inproceedings{radford2023robust,
  title={Robust speech recognition via large-scale weak supervision},
  author={Radford, Alec and Kim, Jong Wook and Xu, Tao and Brockman, Greg and McLeavey, Christine and Sutskever, Ilya},
  booktitle={International conference on machine learning},
  pages={28492--28518},
  year={2023},
  organization={PMLR}
}

@article{xu2025qwen2,
  title={Qwen2. 5-omni technical report},
  author={Xu, Jin and Guo, Zhifang and He, Jinzheng and Hu, Hangrui and He, Ting and Bai, Shuai and Chen, Keqin and Wang, Jialin and Fan, Yang and Dang, Kai and others},
  journal={arXiv preprint arXiv:2503.20215},
  year={2025}
}

@article{hu2022lora,
  title={Lora: Low-rank adaptation of large language models.},
  author={Hu, Edward J and Shen, Yelong and Wallis, Phillip and Allen-Zhu, Zeyuan and Li, Yuanzhi and Wang, Shean and Wang, Lu and Chen, Weizhu and others},
  journal={ICLR},
  volume={1},
  number={2},
  pages={3},
  year={2022}
}

@article{zhang2024mm,
  title={Mm-llms: Recent advances in multimodal large language models},
  author={Zhang, Duzhen and Yu, Yahan and Dong, Jiahua and Li, Chenxing and Su, Dan and Chu, Chenhui and Yu, Dong},
  journal={arXiv preprint arXiv:2401.13601},
  year={2024}
}

@article{caffagni2024revolution,
  title={The revolution of multimodal large language models: a survey},
  author={Caffagni, Davide and Cocchi, Federico and Barsellotti, Luca and Moratelli, Nicholas and Sarto, Sara and Baraldi, Lorenzo and Cornia, Marcella and Cucchiara, Rita},
  journal={arXiv preprint arXiv:2402.12451},
  year={2024}
}

@inproceedings{poswiata-perelkiewicz-2022-opi,
    title = "{OPI}@{LT}-{EDI}-{ACL}2022: Detecting Signs of Depression from Social Media Text using {R}o{BERT}a Pre-trained Language Models",
    author = "Po{\'s}wiata, Rafa{\l}  and
      Pere{\l}kiewicz, Micha{\l}",
    editor = "Chakravarthi, Bharathi Raja  and
      Bharathi, B  and
      McCrae, John P  and
      Zarrouk, Manel  and
      Bali, Kalika  and
      Buitelaar, Paul",
    booktitle = "Proceedings of the Second Workshop on Language Technology for Equality, Diversity and Inclusion",
    month = may,
    year = "2022",
    address = "Dublin, Ireland",
    publisher = "Association for Computational Linguistics",
    url = "https://aclanthology.org/2022.ltedi-1.40/",
    doi = "10.18653/v1/2022.ltedi-1.40",
    pages = "276--282"
}

@article{touvron2023llama,
  title={Llama 2: Open foundation and fine-tuned chat models},
  author={Touvron, Hugo and Martin, Louis and Stone, Kevin and Albert, Peter and Almahairi, Amjad and Babaei, Yasmine and Bashlykov, Nikolay and Batra, Soumya and Bhargava, Prajjwal and Bhosale, Shruti and others},
  journal={arXiv preprint arXiv:2307.09288},
  year={2023}
}

@article{yang2024qwen2,
  title={Qwen2 technical report},
  author={Yang, An and Yang, Baosong and Hui, Binyuan and Zheng, Bo and Yu, Bowen and Zhou, Chang and Li, Chengping and Li, Chengyuan and Liu, Dayiheng and Huang, Fei and Dong, Guanting and others},
  journal={arXiv preprint arXiv:2407.10671},
  year={2024}
}

@article{chen2023speechformer++,
  title={Speechformer++: A hierarchical efficient framework for paralinguistic speech processing},
  author={Chen, Weidong and Xing, Xiaofen and Xu, Xiangmin and Pang, Jianxin and Du, Lan},
  journal={IEEE/ACM Transactions on Audio, Speech, and Language Processing},
  volume={31},
  pages={775--788},
  year={2023},
  publisher={IEEE}
}

@article{haque2018measuring,
  title={Measuring depression symptom severity from spoken language and 3D facial expressions},
  author={Haque, Albert and Guo, Michelle and Miner, Adam S and Fei-Fei, Li},
  journal={arXiv preprint arXiv:1811.08592},
  year={2018}
}

@article{loshchilov2017decoupled,
  title={Decoupled weight decay regularization},
  author={Loshchilov, Ilya and Hutter, Frank},
  journal={arXiv preprint arXiv:1711.05101},
  year={2017}
}

@article{dettmers2023qlora,
  title={Qlora: Efficient finetuning of quantized llms},
  author={Dettmers, Tim and Pagnoni, Artidoro and Holtzman, Ari and Zettlemoyer, Luke},
  journal={Advances in neural information processing systems},
  volume={36},
  pages={10088--10115},
  year={2023}
}

@article{guo2022topic,
  title={A topic-attentive transformer-based model for multimodal depression detection},
  author={Guo, Yanrong and Zhu, Chenyang and Hao, Shijie and Hong, Richang},
  journal={arXiv preprint arXiv:2206.13256},
  year={2022}
}

@inproceedings{toto2021audibert,
  title={Audibert: A deep transfer learning multimodal classification framework for depression screening},
  author={Toto, Ermal and Tlachac, ML and Rundensteiner, Elke A},
  booktitle={Proceedings of the 30th ACM international conference on information \& knowledge management},
  pages={4145--4154},
  year={2021}
}

@inproceedings{jung2024hique,
  title={Hique: Hierarchical question embedding network for multimodal depression detection},
  author={Jung, Juho and Kang, Chaewon and Yoon, Jeewoo and Kim, Seungbae and Han, Jinyoung},
  booktitle={Proceedings of the 33rd ACM International Conference on Information and Knowledge Management},
  pages={1049--1059},
  year={2024}
}

@article{campbell2022speech,
  title={Speech and text processing for major depressive disorder detection},
  author={Campbell, Edward L and Doc{\i}o-Fern{\'a}ndez, Laura and Cummins, Nicholas and Garc{\i}a-Mateo, Carmen},
  journal={Training},
  volume={31},
  pages={76},
  year={2022}
}

@inproceedings{al2018detecting,
  title={Detecting depression with audio/text sequence modeling of interviewsƒ.},
  author={Al Hanai, Tuka and Ghassemi, Mohammad M and Glass, James R},
  booktitle={Interspeech},
  pages={1716--1720},
  year={2018}
}

@inproceedings{wei2022multi,
  title={Multi-modal depression estimation based on sub-attentional fusion},
  author={Wei, Ping-Cheng and Peng, Kunyu and Roitberg, Alina and Yang, Kailun and Zhang, Jiaming and Stiefelhagen, Rainer},
  booktitle={European Conference on Computer Vision},
  pages={623--639},
  year={2022},
  organization={Springer}
}

@inproceedings{williamson2016detecting,
  title={Detecting depression using vocal, facial and semantic communication cues},
  author={Williamson, James R and Godoy, Elizabeth and Cha, Miriam and Schwarzentruber, Adrianne and Khorrami, Pooya and Gwon, Youngjune and Kung, Hsiang-Tsung and Dagli, Charlie and Quatieri, Thomas F},
  booktitle={Proceedings of the 6th international workshop on audio/visual emotion challenge},
  pages={11--18},
  year={2016}
}

@inproceedings{othmani2021towards,
  title={Towards robust deep neural networks for affect and depression recognition from speech},
  author={Othmani, Alice and Kadoch, Daoud and Bentounes, Kamil and Rejaibi, Emna and Alfred, Romain and Hadid, Abdenour},
  booktitle={International conference on pattern recognition},
  pages={5--19},
  year={2021},
  organization={Springer}
}

\appendix

\section{Implementation Details}
\label{appendix:implementation_details}
Our framework is implemented using the HuggingFace \textit{transformers} library with PyTorch 2.1. The full hyperparameter configurations used during training are summarized in Table~\ref{table:implementation_details}. We adopt the AdamW optimizer \citep{loshchilov2017decoupled} for model optimization. To improve training speed without compromising performance, we enable TensorFloat32 (TF32) computation and apply automatic mixed-precision training using BFloat16 (BF16). For parameter-efficient fine-tuning (PEFT) of the Qwen2-Audio model on the depression detection task, we employ QLoRA \citep{dettmers2023qlora}, which compresses the base model to 4-bit precision to reduce memory usage and improve computational efficiency. The full training process requires approximately 90+ GPU hours on an NVIDIA H200 141GB GPU. This includes around 40 hours for self-supervised visual pretraining, 20 hours for utterance-level audio-visual alignment, and 30 hours for multimodal instruction tuning. Early stopping is applied in all stages when training loss plateaus.

\begin{table}[ht]
\footnotesize
\centering
\begin{tabular}{cccc}
\toprule
\multicolumn{1}{c}{}        & Stage I & Stage II    & Stage III   \\ 
\midrule
Optimizer                   & \multicolumn{3}{c}{AdamW}           \\
Learning Rate               & 1.5e-4  & 1e-6        & 3e-6        \\
$\beta_1$                   & \multicolumn{3}{c}{0.9}             \\
$\beta_2$                   & 0.95    & \multicolumn{2}{c}{0.999} \\
Weight Decay                & 0       & \multicolumn{2}{c}{0.001} \\
Batch Size                  & 128     & 64          & 8           \\
\makecell{Grad Accum Steps} & 8       & 16          & 8           \\
Scheduler                   & \multicolumn{3}{c}{Cosine LR}           \\
Num Epochs                  & 50      & 20          & 3           \\
Warm Up Epochs              & 5       & 2           & 0.1         \\
Max Grad Norm               & 1.0     & \multicolumn{2}{c}{0.5}   \\
BF16                        & \multicolumn{3}{c}{True}            \\
TF32                        & \multicolumn{3}{c}{True}            \\ 
\bottomrule
\end{tabular}
\caption{Training hyperparameters.}
\label{table:implementation_details}
\end{table}
\begin{algorithm}[ht]
\centering
\footnotesize
\caption{Time-Sync Data Augmentation}
\begin{algorithmic}[1]
\STATE $N^+ \leftarrow$ Number of positive samples in the training set
\STATE $N^- \leftarrow$ Number of negative samples in the training set
\STATE Set number of subdialogues per positive sample $M^+$
\STATE Set minimum length of subdialogue in seconds $d_{min}$
\STATE Set maximum length of subdialogue in seconds $d_{max}$
\STATE $M^- = N^- / N^+ \times M^+ \leftarrow$ Number of sub-dialogues per negative sample
\FOR{Dialogue $X^{(n)}=(T^{n}, A^{n}, V^{n}), n=1, 2, ..., N$}
    \STATE $D \leftarrow $ Dialogue length in seconds
    \STATE $\{\mathbf{\varepsilon}_{i}\} \leftarrow$ Interviewer utterance start timestamps
    \STATE $\{\mathbf{\varepsilon}_{p}\} \leftarrow$ Participant utterance end timestamps
    \IF{$X^{(n)}$ is positive}
        \STATE $M \leftarrow M^+$
    \ELSE
        \STATE $M \leftarrow M^-$
    \ENDIF
    \FOR{Sub-dialogue $X^{(n)m}$, $m = 1$ to $M$}
        \STATE Sample length $d$ uniformly from $(d_{min}, d_{max})$
        \STATE Sample start timestamp $\varepsilon_s' \in \{\mathbf{\varepsilon}_{i}\}$ from range (0, $D - d$)
        \STATE Round the start timestamp to its closet integer second $\varepsilon_s \leftarrow \lfloor\varepsilon_s'\rfloor$ 
        \STATE $\varepsilon_{tmp} = \varepsilon_s + d \leftarrow$ Raw end timestamp
        \STATE Sample end timestamp $\varepsilon_e' \in \{\mathbf{\varepsilon}_{p}\}$ and $\min |\varepsilon_e' - \varepsilon_{tmp}|$
        \STATE Round the end timestamp to its closet integer second $\varepsilon_e \leftarrow \lceil \varepsilon_e' \rceil$ 
        \STATE Generate subdialogue $T^{(n)m} \leftarrow T^{(n)}_{\varepsilon_s:\varepsilon_e}$
        \STATE Obtain the raw audio segment $A'^{(n)m} \leftarrow A^{(n)}_{\varepsilon_s:\varepsilon_e}$
        \STATE Obtain the raw visual segment $V'^{(n)m} \leftarrow V^{(n)}_{\varepsilon_s:\varepsilon_e}$
        \STATE Remove the interviewer utterances $A^{(n)m} \leftarrow A'^{(n)m}$ and $V^{(n)m} \leftarrow V'^{(n)m}$
        \STATE Subdialogue $X^{(n)m} = (T^{(n)m}, A^{(n)m}, V^{(n)m})$
    \ENDFOR
\ENDFOR
\end{algorithmic}
\end{algorithm}
\section{Details of Timestamp-Synchronized Data Augmentation}
\label{appendix:data_augmentation}
Following the approach of \citet{wu2023self}, we generate subdialogues from the original interview transcripts to mitigate class imbalance and expand the size of the training set. In our data augmentation pipeline, we enforce strict synchronization among transcripts, audio, and video to ensure precise timestamp-level alignment. However, due to varying frame rates across modalities, achieving synchronization presents a technical challenge. For example, audio recordings are typically captured at a 16,000 Hz sampling rate and later converted into Mel-spectrograms with a frame rate of 100 Hz, while video recordings are collected at 30 frames per second (FPS). To address this discrepancy, we constrain the start and end timestamps of each subdialogue to align with whole seconds (i.e., integer-second boundaries). 

Additionally, we require each subdialogue to begin with an utterance from the interviewer and conclude with a response from the participant. This design choice ensures that each subdialogue forms a complete and contextually coherent conversational unit, with a clear initiation and response structure. Such a constraint preserves the semantic continuity and logical flow within each segment, making them more suitable for downstream tasks that rely on natural discourse patterns. Moreover, this structure aligns with the training paradigm of large language models, which are typically pretrained on large-scale dialogue corpora. By maintaining this dialogue consistency, we enhance the model’s ability to interpret the subdialogues effectively within a familiar conversational framework.

\section{The Prompt Design for Instruction Tuning}
\label{appendix:tuning}
During instruction tuning, we design a system prompt to guide the behavior of the language model. Given that the Qwen2-Audio-Instruct model has been fine-tuned on audio analysis tasks, we adopt a chat-based prompt template to elicit model responses. Notably, we use the same prompt design for both the Qwen2-Audio family and our multi-modal LLM. This consistency is based on our integration strategy, where visual embeddings are directly added to the audio embeddings without modifying the model architecture. Therefore, we assume that the model can still function effectively even without explicitly referencing visual information in the prompt.
\paragraph{System Prompt}
\texttt{Below is a conversation between an interviewer and a participant. Please analyze the transcripts and audio, and find whether the participant is affected by depression.}
\paragraph{Instructions}
\texttt{Audio: \{audio\} \textbackslash n Interview conversation: \{transcripts\} \textbackslash n Response: \textbackslash n}

\end{document}